\documentclass[12pt,preprint]{aastex}

\newcommand\percc  {\hbox{cm$^{-3}$}}                  % cm^-3
\newcommand\kms    {\hbox{km{\hskip0.1em}s$^{-1}$}}    % km/s
\newcommand\etal   {{\it et al. }}                     % et al.
\newcommand\amin   {$^{\prime}$}                       % Arcminutes symbol
\newcommand\damin  {\hbox{$.\mkern-4mu^\prime$}}       % Arcminutes over dot
\begin{document}

\title{Discovery of an OH(1720 MHz) Maser in the LMC}

\author{D.A. Roberts\altaffilmark{1}}
\affil{Department of Physics and Astronomy, Northwestern University,
Evanston, IL. 60208}
\email{doug-roberts@northwestern.edu}

\author{F. Yusef-Zadeh}
\affil{Department of Physics and Astronomy, Northwestern University,
Evanston, IL. 60208}
\email{zadeh@northwestern.edu}

\altaffiltext{1}{Adler Planetarium \& Astronomy Museum,
1300 S. Lake Shore Drive, Chicago, IL. 60605}

\begin{abstract}

We report the first study of OH(1720 MHz) masers in the LMC in order
to probe regions where supernova remnants interact with adjacent
molecular clouds. Using the ATCA, we observed four sources in the LMC
and detected a single OH(1720 MHz) maser with a flux density of 377
mJy toward 30 Doradus.  No main line emission at 1665 or 1667 MHz was
detected. The observed OH(1720 MHz) maser emission from 30 Dor shows
characteristics similar to the well-known collisionally-pumped
supernova remnant masers found in the Galaxy, though 30 Dor is known
as a star forming region.  It is possible that shocks driven by a
nearby supernova remnant or by strong stellar winds from young stars
are responsible for production of OH(1720 MHz) maser in 30 Dor.
Future studies are required to distinguish between collisional and
radiative pumping mechanisms for the 30 Dor OH (1720 MHz) maser.

\end{abstract}

%\keywords{Galaxies: nuclei --- Galaxies: The Galaxy --- Radio sources: 
%Lines --- interferometry}

\section{Introduction}

In the past decade, OH(1720 MHz) maser emission has been found to be a
powerful tool to investigate how shocks from supernova remnants
interact with molecular clouds in the Galaxy (e.g. Frail \etal 1994;
Wardle \& Yusef-Zadeh 2002). In these cases, the so-called supernova
remnant (SNR) masers are thought to be collisionally excited by the
passage of the supernova remnant shock through the surrounding
molecular clouds.  The physical conditions for the population
inversion and amplification in the 1720 MHz line alone requires
densities $\sim 10^5$ \percc, temperatures in the range 30--125 K, an
OH column density in the range $10^{16}$--$10^{17}$ cm$^{-2}$ and an
absence of a strong FIR continuum (Elitzur 1976; Lockett, Gauthier \&
Elitzur 1999).  The latter constraint explains why this class of
OH(1720 MHz) masers is rare in star forming regions where the UV
radiation from hot, massive young stars is absorbed by dust grains and
re-radiated in the FIR.

The OH molecule has two rotational ladders, $^2\Pi_{3/2}$ and
$^2\Pi_{1/2}$; all levels are further split by lambda doubling and
hyperfine splitting.  Ground-state transitions include the two main
lines of OH, at 1665 and 1667 MHz, and the two satellite lines, at
1612 and 1720 MHz. Radiative pumping can produce strong OH maser
emission at 1665, 1667 and 1612 MHz, but can provide only weak
emission (relative to 1665 and 1667 MHz lines) at 1720 MHz. However,
collisional pumping in SNR masers produces strong maser emission in
the OH(1720 MHz) line, with no emission from other ground state
transitions.

A total of about 20 SNR masers have been found in the Galaxy (e.g.,
Green et al. 1997; Koralesky et al. 1998; Yusef-Zadeh et al. 1999).
Motivated by the bright maser line emission from the Galactic center
and W28, we searched for SNR masers in the 30 Dor region of the LMC.
Here, we present the first discovery of OH(1720 MHz) line emission in
the LMC.  The observed flux density is 377 mJy; if this OH(1720 MHz)
maser were at the distance of the Galactic center, it would be 12 Jy.
This is withing a factor of two of the observed flux density
($\approx$7.4 Jy) of the strongest OH(1720 MHz) maser in Sgr A East,
which is known as a SNR interacting with an adjacent molecular cloud
near the Galactic center (Yusef-Zadeh et al. 1999).

These initial results reported here were first presented in the
workshop ``10$^{51}$ Ergs: The Evolution of Shell Supernova
Remnants,'' hosted by the University of Minnesota, 1997 March 23-26.
Another analysis of this maser line emission from the LMC has
recently been given by Brogan et al. (2004).
 
\section{Observations \& Results}

In an initial observation in 1996, three fields (30 Dor, N44, and N49)
were observed in a single 12 hour period with 6-kilometer array of the
Australia Telescope Compact Array (ATCA) to search for OH(1720 MHz)
emission. In this first observation, a single strong maser was
identified near the core of 30 Dor. In order to confirm the detection
of the 30 Dor maser in 1997 we used the ATCA to conduct a 13-hour
follow-up observation of the initial three fields and an additional
field (N132D) in the OH(1720 MHz) line.  An additional 13 hours was
spent on the four fields to look for OH emission in the main lines at
1665 and 1667 MHz.  These regions were chosen because they have some
of the largest concentration of SNRs and molecular clouds in the
LMC. For all observations, 1934-638 was used as the primary calibrator
and 0407-658 was used to calibrate the bandpass and complex gains. The
important observational parameters are presented in Table
\ref{tab:obsparm}.

All calibration and data reduction were carried out using the
Multichannel Image Reconstruction, Image Analysis and Display
(MIRIAD). The antenna gains and bandpasses were calibrated and then
the continuum emission was subtracted in the visibility domain using
UVLIN program of MIRIAD.  For 30 Dor, the observations from 1996 and
1997 were processed separately in order to look for variability.  In
order to search for maser emission, the entire 4 MHz bandwidth was
imaged using natural weighting across the entire 30\amin\ primary beam
of the ATCA; a cube with 1024 Hanning-smoothed spectral line channels
(with a channel resolution of $\sim$0.68 \kms) was created. In the case
of the 30 Dor field, the two days' data were imaged separately in order
to look for variation in the maser strength. In order to determine the
maser line characteristics, the channels within $\pm$ 30 \kms\ of the
detected 30 Dor maser were imaged at the highest velocity resolution
(i.e., not Hanning smoothed) of 0.34 \kms. All image cubes were
deconvolved using the CLEAN algorithm of MIRIAD. Because the 30 Dor
maser was not at the center of the observed field, the final
deconvolved images were corrected for the attenuation of the primary
beam.

\begin{figure}[htb]
\includegraphics[width=1.0\textwidth, height=1.0\textwidth,
angle=0]{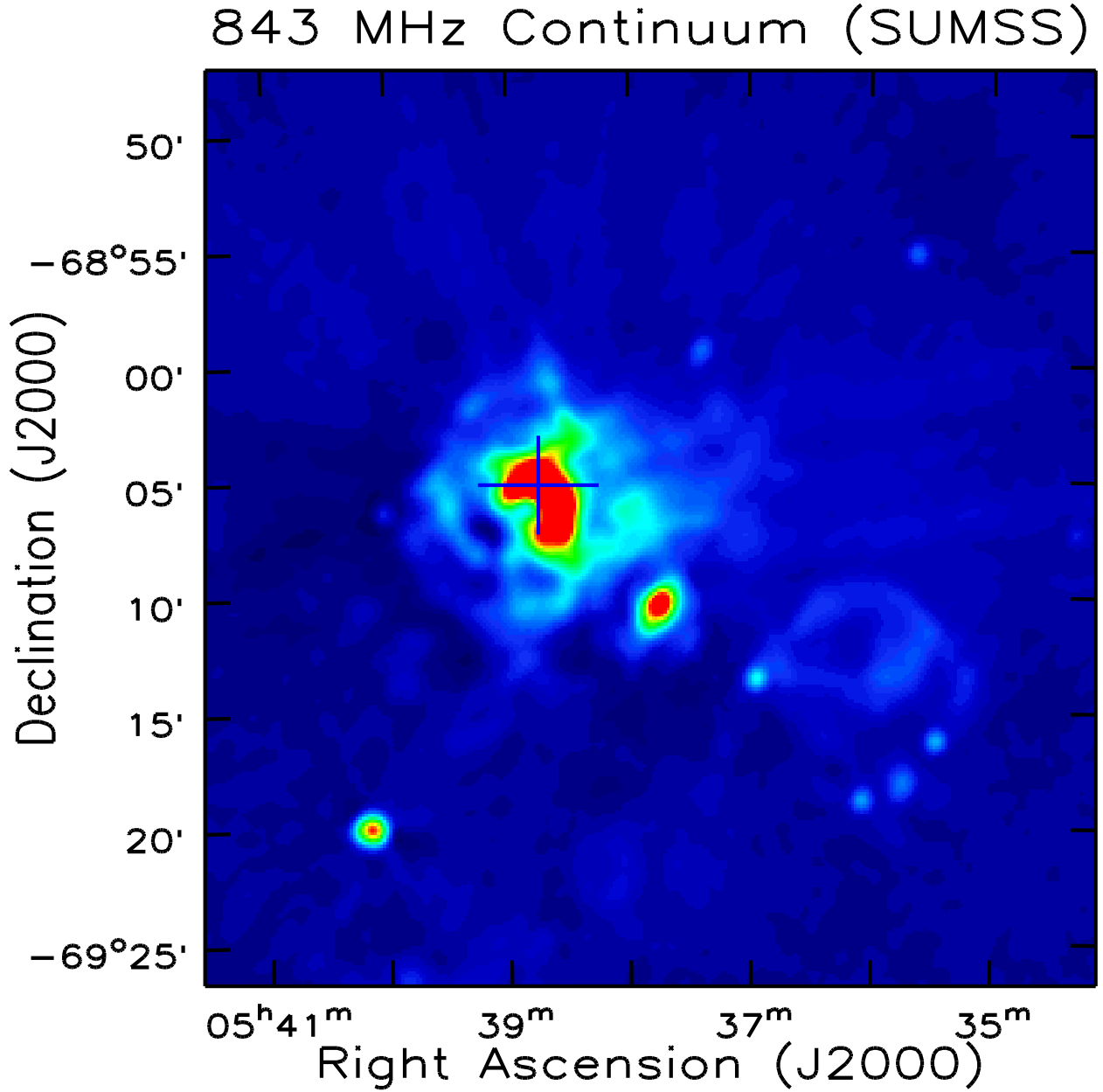} 
\caption{Pseudocolor image of 30 Dor from Molonglo Observatory
Synthesis Telescope (MOST) at 843 MHz from the SUMSS at a resolution
of 11${{\rlap.}^{\prime\prime}}$3 $\times$
8${{\rlap.}^{\prime\prime}}$2. The cross (+) marks the position of the
detected OH(1720 MHz) maser.  The spectra at this position from all
three observed OH lines are displayed in Fig. \ref{fig:spectra}.}
\label{fig:cont}
\end{figure}

\begin{figure}[htb]
\includegraphics[width=1.0\textwidth, height=1.0\textwidth,
angle=0]{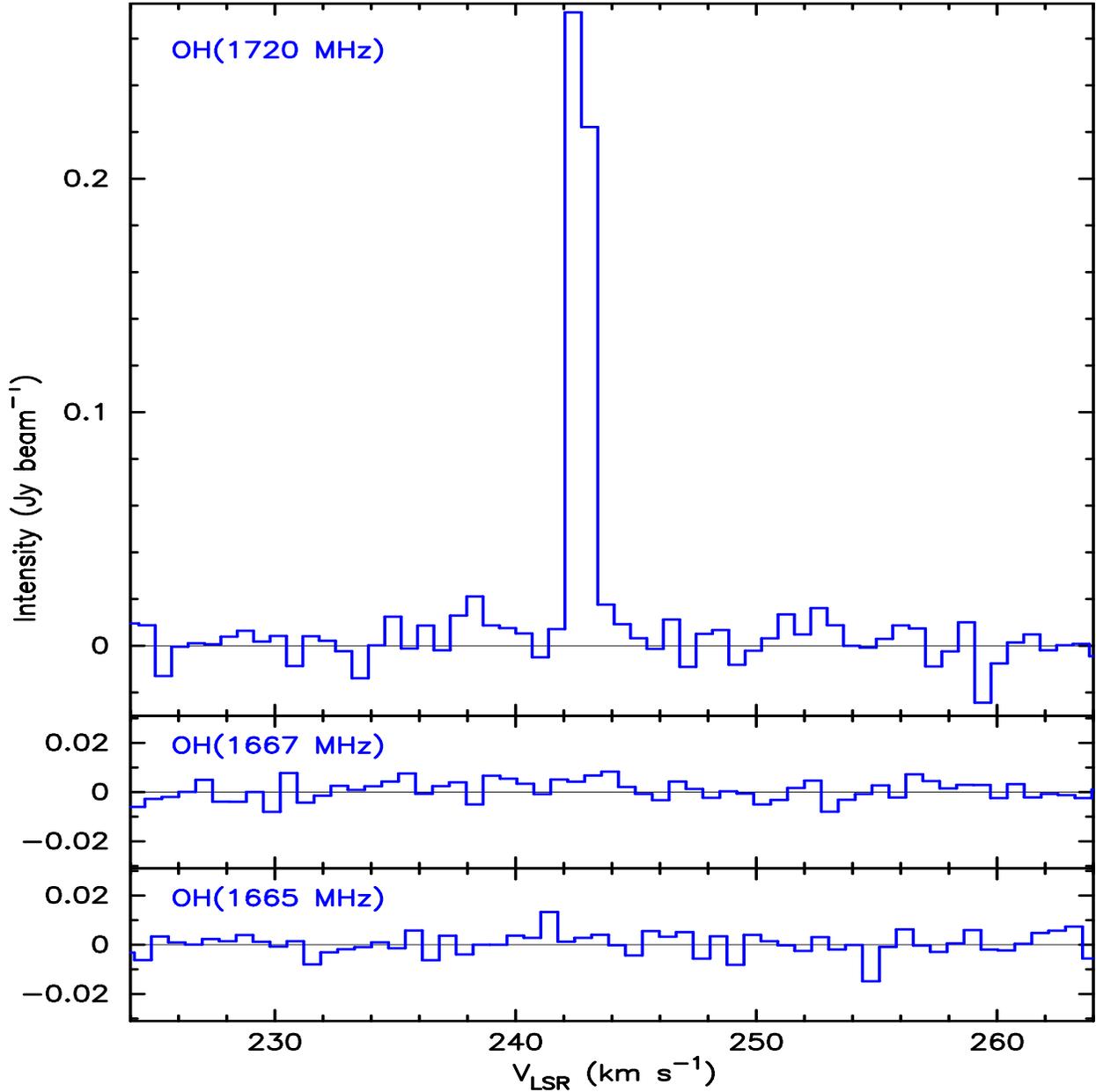} 
\caption{Plots of OH emission (1665, 1667 \& 1720 MHz) at the
location of the detected OH (1720 MHz) maser in 30 Dor
(+ in Fig. \ref{fig:cont}).  The spectral and spatial
parameters of the maser emission are given in Table
\ref{tab:maserparm}.}
\label{fig:spectra}
\end{figure}

Out of the four fields observed, the only emission detected was a
single OH(1720 MHz) maser near the core of 30 Dor. The properties of
the maser are given in Table \ref{tab:maserparm}.  The position was
determined by fitting a two-dimensional Gaussian (using IMFIT in
MIRIAD) to the strongest channel of the data.  The line width and
center velocity were determined by fitting a Gaussian to the line
profile with the PROFIT program of the Groningen Image Processing
SYstem (GIPSY). The flux density of the maser was determined
separately for the two observations in order to determine if any
significant variation had occurred.  A continuum image of the 30 Dor
field from the Sydney University Molonglo Sky Survey (SUMSS, Bock,
Large, \& Sadler 1999) is presented in Fig. \ref{fig:cont}, with the
position of the detected maser shown by the dark cross.  A spectrum of
the OH(1720 MHz) maser in 30 Dor is shown in Fig. \ref{fig:spectra}.
In the OH(1720MHz) line, the only maser detected was the strong one in
30 Dor.  In the OH lines at 1665 and 1667 MHz, no masers were detected
in any field; in particular no main line masers were detected at the
position of the 30 Dor OH(1720 MHz) maser.

The strong OH(1720 MHz) maser detected in the initial observation was
confirmed in the follow-up observation a year later.  The intensity of
the maser did not change significantly between the two observations;
in the initial observation the maser was 337 $\pm$ 34 mJy beam$^{-1}$
and in the follow-up observation it was 370 $\pm$ 28 mJy beam$^{-1}$.
We present 3-$\sigma$ upper limits for all three lines and all four
fields in Table \ref{tab:maserparm}.  

Our results are in general agreement with those reported by Brogan et
al. (2004).  In particular, Brogan et al. report two additional weak
($<$ 28 mJy beam$^{-1}$) masers in the 30 Dor field.  In our
observations, the noise toward these masers was too large (12 mJy
beam$^{-1}$) to have detected them.  The noise of the 30 Dor
observations was larger than that toward the other fields because the
pointing center of the 30 Dor observations was not centered on the
maser position (12\damin8 distant in 1996 observations).  Brogan et
al. also reported the detection of two weak OH(1720 MHz) masers in N49
at the level of 25-35 mJy beam${-1}$.  After careful analysis of our
two OH(1720 MHz) observations toward N49, we do not detect the masers
reported by Brogan et al.  The 1996 and 1997 datasets have
sensitivities (3-$\sigma$ detection thresholds) of 16 and 21 mJy
beam$^{-1}$, respectively.  It is possible that the maser lines in N49
have a line width smaller than our channel width (our channel width is
0.68 km s${-1}$, compared with 0.34 km s${-1}$ for Brogan et al.).  It
is also possible that the maser emission is variable.  An additional
epoch of sensitive observations are needed to confirm the potential
variability.

\section{Discussion}

The coincidence between the OH(1720 MHz) maser emission and the
prominent 30 Dor region raises the obvious question of whether the
maser source is radiatively-pumped in the HII complex region (or the
cluster of young stars R136) or collisionally excited by the passage
of a shock running into a molecular cloud.  The possible
photo-excitation of the OH(1720 MHz) maser by the HII region in 30 Dor
can be investigated by observing the main OH lines at 1665 and 1667
MHz.  In a previous observation of 30 Dor with the Parkes telescope
having 12\damin5 resolution, thermal absorption is detected at a level
of 90 and 140 mJy at the LSR velocities of 245 and 247 \kms,
respectively with no evidence of emission (Gardner \& Whiteoak, 1985).
This is consistent with our results from our ATCA observations at 1665
and 1667 MHz, in which no emission was found (see Table
\ref{tab:maserparm}). Additionally, no absorption was detected in our
observation, suggesting that the thermal absorption observed at Parkes
was resolved out by the ATCA (which have a minimum spacing of 214 m,
corresponding to a limit to the largest observable structure of
1\damin6).  In total, the OH results from our ATCA observations and
those from Parkes show two components: smooth thermal absorption in
all the OH ground-state transitions at 1612, 1665, 1667, and 1720 MHz
along with a spatially-compact, spectrally-narrow maser line at only
detected at 1720 MHz. The OH properties are consistent with a model,
in which diffuse cold molecular gas, which is observed in OH
absorption, exists around the HII regions, and a shock driving into
the molecular gas has dissociated water in the gas, increasing the OH
abundance and exciting the OH(1720 MHz) line.

SNR masers are thought to be collisionally excited by C-type shocks
passing through a molecular cloud.  These shocks are usually detected
at the interface of SNRs and adjacent molecular clouds.  It is
possible that the OH(1720 MHz) maser in the LMC may be associated with
the shock from an obscured supernova in the vicinity of the HII
region.  The nebula is characterized by filamentary optical structures
filled with X-ray emitting gas. It has been speculated that supernova
shocks may play a key role in the formation of the nebula and in the
heating of the hot gas (Bamba et al. 2003).  Indeed, a number of young
SNRs including N157B, have been identified in the vicinity of 30 Dor
(Chu et al. 1995; Lazendic, Dickel \& Jones 2003). Furthermore, two
high-mass X-ray binary candidates have been discovered in the core of
the nebula, suggesting recent supernova explosions at or near the
central star cluster R136 (Wang 1995).  It is also possible that the
maser could be pumped by collisions between strong stellar winds into
a clumpy environment.  Portegies Zwart, et al. (2002) have determined
that the X-ray point-source emission observed from Chandra can be
explained by a colliding wind binaries.  Shocks from winds into the
dense molecular environment could result in the observed OH(1720 MHz)
emission.  There is clear evidence of molecular gas in the 30 Dor
region as indicated by the detection of CO emission using the SEST
mm-wave telescope (Johansson et al. 2003) and of extended 2.1212
$\mu$m H$_2$ 1--0 $S(1)$ ro-vibrational line emission using the 2.2 m
ESO-MPIA telescope in La Silla with the FAST imaging NIR-spectrometer
(Krabbe et al. 1991).  However, firm identifications of SNRs and
molecular clouds within the nebula have been difficult, primarily
because of the overwhelming thermal radio continuum and optical
background. It has been suggested that many SNRs may hide inside the
nebula.

Most observations of star forming regions show ground-state OH
(1665/67 MHz) masers which are thought to be pumped radiatively.  The
unaccompanied OH(1720 MHz) SNR masers are believed to be
collisionally excited.  This distinction is under refinement due to
recent observations of the $^2\Pi_{1/2}$, {\sl J}=1/2 excited state of
OH at 4765 MHz which is detected in a number of star forming regions
where OH(1720 MHz) maser line emission has also been detected
(Palmer, Goss \& Whiteoak 2004; Palmer, Goss \& Divine 2003).
Moreover, in a number of star forming regions excited OH(4765 MHz)
masers are spatially coincident with OH(1720 MHz) ones.  In W3(OH),
measurement of these coincidences has been most precise so far: about
one-third of the 4765 MHz maser spots coincide with 1720 MHz spots to
an accuracy $<$ 10 AU (Palmer, Goss \& Whiteoak 2004).  This spatial
correlation suggests that both the excited OH (4765 MHz) maser
emission and ground state OH(1720 MHz) masers can be pumped
radiatively in environments of high densities and temperatures (Gray
et al. 2001).  It will be important to investigate whether OH (4765
MHz) emission is coincident with the detected OH(1720 MHz) masers in
30 Dor.  If the two maser emissions are correlated in 30 Dor, it would
suggest a radiative pumping mechanism. However, the lack of main line 
OH emission at 1665/67 MHz is puzzling in this radiative pumping scheme. 
Thus, we believe that OH(1720 MHz) maser line emission is  collisionally
pumped, either by SNR or wind interactions with surrounding molecular
clouds. 

\section{Summary}

We report the detection of the first OH(1720 MHz) maser in the
LMC. Out of four fields (30 Dor, N44, N49, and N132D) a single strong
OH(1720 MHz) maser is detected near the center of 30 Doradus. The
maser emission was confirmed by subsequent observations and no
significant variation was detected in the 11 months between the two
observations.  No significant emission was detected in the OH 1665 and
1667 MHz lines toward the OH(1720 MHz) maser or in any of the observed
fields. The lack of OH(1665 and 1667MHz) maser emission coincident
with the OH(1720 MHz) maser strongly suggests that the 1720 MHz maser
is collisionally-excited, rather than radiatively-pumped from the
nearly 30 Dor HII regions.

\acknowledgements{DAR thanks the National Radio Astronomy Observatory
and its financial support for U.S. astronomers observing with foreign
telescopes.}

\begin{deluxetable}{ll}
\tablenum{1}
\tablecolumns{2}
\tablewidth{0pt}
\tablecaption{Observational Parameters}
\tablehead{
\colhead{\bf Observing Parameter} &
\colhead{\bf Value}
}
\startdata
ATCA Array & 6 km configuration\\
Velocity coverage~~~~ & --83 to +614 km s$^{-1}$\\
Velocity resolution~~~~ & 0.68 km s$^{-1}$ (Hanning smoothed)\\
Resolution & 8$^{\prime\prime} \times 7^{\prime\prime}$\\
Field of view & 30$^\prime$\\
\hline
{\bf Initial} & \\
~~Date & July 19, 1996 \\
~~Transitions & OH(1720 MHz)~~\\
~~Fields & 30 Dor, N44 \& N49\\
~~Total observing time & 2 $\times$ 8 hr \\
\hline
{\bf Follow-up} & \\
~~Date & May 9, 1997 \\
~~Transitions & OH 1665, 1667, 1720 MHz~~\\
~~Fields & 30 Dor, N44, N49, N132D\\
~~Total observing time & 1.4 $\times$ 8 hr \\
\enddata
\label{tab:obsparm}
\end{deluxetable}

\begin{deluxetable}{ll}
\tablenum{2}
\tablecolumns{2}
\tablewidth{0pt}
\tablecaption{Maser Parameters}
\tablehead{
\colhead{\bf 30 Dor Field} & 
}
\startdata
OH(1665 MHz) & 18.9 mJy beam$^{-1}$ (3-$\sigma$ upper limit)\\
OH(1667 MHz) & 21.1 mJy beam$^{-1}$ (3-$\sigma$ upper limit)\\
OH(1720 MHz) & {\bf Initial -- July 1996}\\
~~~$\bullet$~Line Peak & 377 $\pm$ 34 mJy beam$^{-1}$ \\
~~~$\bullet$~Line FWHM & 0.861 $\pm$ 0.087 km s$^{-1}$ \\
~~~$\bullet$~{\sl V}$_{\rm LSR}$ & 242.693 $\pm$ 0.012 km s$^{-1}$ \\
~~~$\bullet$~Position & \\
~~~~~~~RA(J2000) & 05$^{\rm h}$ 38$^{\rm h}$ 45${{\rlap.}^{\rm s}}$00\\
~~~~~~~Dec(J2000) & --69$^{\circ}$ 05$^{\prime}$ 07${{\rlap.}^{\prime\prime}}$5\\
~~~$\bullet$~Extent & 4${{\rlap.}^{\prime\prime}}$5 $\times$
0${{\rlap.}^{\prime\prime}}$6, PA=112$^\circ$\\
OH(1720 MHz) & {\bf Follow-up -- May 1997}\\
~~~$\bullet$~Line Peak & 370 $\pm$ 28 mJy beam$^{-1}$ \\
~~~$\bullet$~Line FWHM & 0.796 $\pm$ 0.068 km s$^{-1}$ \\
~~~$\bullet$~{\sl V}$_{\rm LSR}$ & 242.779 $\pm$ 0.026 km s$^{-1}$ \\
%~~~$\bullet$~Position & (same as initial)\\
\hline
{\bf N44 Field} & \\
\hline
OH(1665 MHz) & 14.4 mJy beam$^{-1}$ (3-$\sigma$ upper limit)\\
OH(1667 MHz) & 17.4 mJy beam$^{-1}$ (3-$\sigma$ upper limit)\\
OH(1720 MHz) & 14.0 mJy beam$^{-1}$ (3-$\sigma$ upper limit)\\
\hline
{\bf N49 Field} & \\
\hline
OH(1665 MHz) & 15.0 mJy beam$^{-1}$ (3-$\sigma$ upper limit)\\
OH(1667 MHz) & 17.1 mJy beam$^{-1}$ (3-$\sigma$ upper limit)\\
OH(1720 MHz) & 15.9 mJy beam$^{-1}$ (3-$\sigma$ upper limit)\\
\hline
{\bf N132D Field} & \\
\hline
OH(1665 MHz) & 14.7 mJy beam$^{-1}$ (3-$\sigma$ upper limit)\\
%OH(1667 MHz) & BAD mJy beam$^{-1}$ (3-$\sigma$ upper limit)\\
OH(1720 MHz) & 21.0 mJy beam$^{-1}$ (3-$\sigma$ upper limit)\\
\enddata
\label{tab:maserparm}
\end{deluxetable}

\begin{thebibliography}{}

\bibitem[Bamba, (2003)]{bamba03}Bamba, A., Masaru, U., Nakajima, H. \&
  Koyama, K. 2003 ApJ 602, 257
\bibitem[Bock,(1999)]{bock99}Bock, D.C.-J, Large, M.I. \& Sadler,
  E.M. 1999, AJ 117, 1578B
\bibitem[Brogan, (2004)]{brogan04}Brogan, C.L., Goss, W.M., Lazendic,
  J.S., \& Green, A.J. 2004 AJ 128, 700
%\bibitem[Caswell, (2004)]{caswell04}Caswell, J.L. 2004, MNRAS 349, 99
\bibitem[Chu, (1995)]{chu95}Chu, Y.-H., Dickel, J.R., Staveley-Smith,
L., Osterberg, J., \& Smith, R. C. 1995, AJ 109, 1729
\bibitem[Elitzur, (1976)]{elitzur76}Elitzur, M. 1976, ApJ 203, 124
\bibitem[Frail, (1994)]{frail94}Frail, D.A., Goss, M.W. \& Slysh,
  V.I. 1994, ApJ 424, L111 
\bibitem[Gardner, (1985)]{gardner85}Gardner, F.F. \& Whiteoak,
  J.B. 1985, MNRAS 215, 103
\bibitem[Gray, (2001)]{gray01}Gray, M.D., Cohen, R.J., Richards,
  A.M.S., Yates, J.A. \& Field, D. 2001, MNRAS 324, 643
\bibitem[Green, (1997)]{green97}Green A.J., Frail, D.A., Goss, W.M. \&
  Otrupcek, R. 1997, AJ 114, 2058
\bibitem[Johansson, (2003)]{johansson03}Johansson, L.E.B., Greve, A.,
  Booth, R.S., Boulanger, F., Garay, G., de Graauw, Th., Israel, F.P.,
  Kunter, M.L., Lequeux, J. Murphy, D.C., Nyman, L.-\AA. \& Rubio,
  M. 1998, A\&A 331, 857
\bibitem[Koralesky, (1998)]{koralesky98}Koralesky, B., Frail, D.A.,
  Goss, W.M., Claussen, M.J. \& Green, A.J. 1998, AJ, 116, 1323
\bibitem[Krabbe, (1991)]{krabbe91}Krabbe, A., Storey, J., Rotaciuc,
  V., Drapatz, S. \& Genzel, R. 1991, in The Magellanic Clouds:
  Proceedings of the 148th Symposium of the IAU, held in Sydney,
  Australia, July 9-13, 1990. Eds.Raymond Haynes \& Douglas Milne. IAU
  Symp. 148, (Kluwer: Dordrecht), 205 
\bibitem[Lazendic, (2003)]{lazendic03}Lazendic, J.S., Dickel, J.R., \&
  Jones, P.A. 2003, ApJ 596, 287
\bibitem[Lockett, (1999)]{locket99}Lockett, P., Gauthier, E. \&
  Elitzur, M. 1999, ApJ 511, 235
\bibitem[Palmer, (2003)]{palmer03}Palmer, P., Goss, W.M. \& Devine,
  K.E. 2003, ApJ, 599, 324
\bibitem[Palmer, (2004)]{palmer04}Palmer, P., Goss, W.M. \& Whiteoak,
  J.B.  2004, MNRAS 347, 1164
\bibitem[Portegies, (2002)]{portegies02} Portegies Zwart, S.F.,
  Pooley, D. \& Lewin, W.H.G. 2002, ApJ 574, 762
\bibitem[Wang, (1995)]{wang95}Wang Q.D. 1995, ApJ 453, 783
\bibitem[Wardle, (2002)]{wardle02}Wardle, M. \& Yusef-Zadeh, F. 2002,
  Science, 296, 2350
\bibitem[Yusef-Zadeh, (1999)]{zadeh99}Yusef-Zadeh, F., Roberts, D.A.,
  Goss, W.M., Frail, D.A. \& Green, A. 1999, ApJ 527, 172

\end{thebibliography}
\end{document}